# Facing Identity:
## The Formation and Performance of Identity via Face-Based Artificial Intelligence Technologies

## Wells Lucas Santo

*Submitted in partial fulfillment of
the Field Preliminary Milestone at
the University of Michigan School of Information*

Oral Defense Date: October 11, 2024

**Committee Members:**
*(in alphabetical order)*

Matthew Bui (co-advisor)
Oliver Haimson
Lisa Nakamura
Sarita Schoenebeck (co-advisor)

# Abstract

How is identity constructed and performed in the digital via face-based artificial intelligence technologies? While questions of identity on the textual Internet have been thoroughly explored, the Internet has progressed to a multimedia form that not only centers the visual, but specifically the face. At the same time, a wealth of scholarship has and continues to center the topics of surveillance and control through facial recognition technologies (FRTs), which have extended the logics of the racist pseudoscience of physiognomy. Much less work has been devoted to understanding how such face-based artificial intelligence technologies have influenced the formation and performance of identity. This literature review considers how such technologies interact with *faciality*, which entails the construction of what a face may represent or signify, along axes of identity such as race, gender, and sexuality. In grappling with recent advances in AI such as image generation and deepfakes, I propose that we are now in an era of "post-facial" technologies that build off our existing culture of facility while eschewing the analog face, complicating our relationship with identity vis-á-vis the face. Drawing from previous frameworks of identity play in the digital, as well as trans practices that have historically played with or *trans*gressed the boundaries of identity classification, we can develop concepts adequate for analyzing digital faciality and identity given the current landscape of post-facial artificial intelligence technologies that allow users to inter*face* with the digital in an entirely novel manner. To ground this framework of *trans*gression, I conclude by proposing an interview study with VTubers — online streamers who perform using motion-captured avatars instead of their real-life faces — to gain qualitative insight on the experience and perceptions of users of post-facial technologies and how these sociotechnical experiences interface with our relationships with identity and the digital anew.

# 1 Introduction

> *"Any system for representing the face tells us something about the society and historical moment that produced it."* – Kelly Gates, *Our Biometric Future* [16]

What's in a face?

In *Digitizing Race*, Lisa Nakamura [37] introduces the case study of the early 2000's website, alllooksame.com, which asks its visitors to attempt to distinguish Chinese, Japanese, and Korean individuals by pictures of their faces. As a trans non-binary Asian American, I grew up hearing various conversations about how one could infer identity characteristics about a person — such as their race or sexuality — just by looking at their facial features. Such discourse often involved invoking stereotypes about essential differences between the various Asian ethnicities, which alllooksame.com actually worked to counter by showing how difficult and unfounded determining ethnicity from the face was and providing a discussion board for visitors to share their experiences with the task. By encouraging such discourse, this website provided an "apparatus that deconstructs the visual culture of race" [37]. Yet, such sentiments have continued to exist. In the late-2010's, advances in artificial intelligence led to researchers proposing that algorithms in the subfield of computer vision could be developed to infer identity characteristics about an individual, purely based on an image of their face. This included projects based in the hope that such technologies could help people in the United States distinguish Asian ethnicities, which are often mistaken for one another, as exemplified in the introduction to one study developed by Stanford computer science students:

> "We can often guess a person's ethnicity by the way he or she looks. However, Chinese, Japanese and Korean have very similar facial anatomy partially due to their close geographical relationships. Often times people from these three countries encounter situations where others mistaken their nationality the first time they meet. This is very common in the United States … We implemented the machine learning approach to classify and predict Chinese, Japanese and Korean based on the facial images." [13]

Many computer vision projects have been employed to determine not only the race and ethnicity of an individual, but what gender they are, and even whether they are queer or not [59]. This has marked a return to the racist pseudoscience of physiognomy within the field of artificial intelligence, where the idea of being able to infer one's innate characteristics purely from their facial features abounds. While much scholarship has focused on the clear and present dangers of surveillance and the targeting of historically oppressed peoples using face-based AI technology, how these technologies have affected our perceptions of what the face signifies in relation to our conceptions of identity have been far less explored.

As critical scholars have argued, digitization and algorithms have helped to usher in a new era of presumed "objectivity," with digital technologies recapturing and reproducing old logics of biological essentialism, convincing us that what face-based algorithms infer is absolute truth, such as what categories of race and gender are "real" [1, 52, 55]. These logics continue to inform the latest advances in artificial intelligence, such as generative AI which can be used to produce images of fake faces and deepfakes which can be used to assume different identities through the changing or masking of the face. These emerging "post-facial" technologies present new terrains to question what beliefs about identity we embed into the face. With these new terrains come new questions, such as what we mean when we ask DALL-E to generate the face of an "Asian" individual, or how online streamers who perform using AI-powered motion-capture technologies think about the face in relation to their performance of a certain identity, that require updated conceptual frameworks to tackle.

In order to address the question of how identity is constructed and performed in the digital via face-based artificial intelligence technologies, this literature review gathers and synthesizes scholarship about digital identity and face-based technologies to build the foundation for the conceptual tools and frameworks to think about identity in this new "post-facial" era. Early work on identity performance, play, and formation in the digital aids us in grounding our analyses, as well as the work from artists and scholars in the space of queer/trans studies who have historically subverted identity through the face. I argue that there is a gap in our scholarship for thinking about the impact of generative AI and other post-facial technologies from a more critical and sociocultural perspective. This topic area is crucial for further exploration, as these technologies are changing the way we interface with the digital and with our sense of identity as expressed through the face. This paper concludes by proposing a qualitative study on VTubers — online streamers who perform using motion-captured avatars instead of their real-life face — as a particularly insightful starting point to begin our explorations into the post-facial.

## 1.1 Scope of Literature Review

There is a plethora of scholarship on facial recognition technologies, especially along the lines of surveillance, fairness, and bias, from computer scientists and critical computing researchers. Though some of the foundational works in this space will be briefly discussed, this literature review focuses moreso on the questions of identity as it relates to the digital face. In order to do so, this literature review touches on early work on identity formation, performance, and play in the digital, from prominent scholars who have studied identity in the textual era of the "cyberspace," such as Sherry Turkle, Lori Kendall, and Lisa Nakamura, which forms the foundation for analyzing how identity can be played with by face-based AI technologies. I then follow Nakamura in shifting to a focus on the visual — specifically, I turn to the facial, drawing from cultural and communication studies analyses of the facial that are rooted in Deleuze and Guattari's conception of *faciality*, without deviating too deeply into a meditation on cultural theory. In analyzing the specific face-based AI technologies that interact with identity, I do not focus on the more "technical" computer science papers about the quantitative improvement of algorithms, but rather look to sociotechnical works that analyze and/or critique the very logics of face-based technologies as they are rooted in the racist pseudoscience of physiognomy and the political project of classification and categorization. In order to scope this paper strictly to the question of how identity (primarily in terms of gender, race, and sexuality) has been constructed and performed through the face, related topics such as emotion recognition and labor, online dating, and the development of racialized cybercultures in online communities, are not covered. Finally, I draw from the work of queer/trans artists and researchers for a crucial body of work on trans technologies and subverting, glitching, or "transgressing" identity through the face, which I use as a framework through which I analyze "post-facial" technologies such as generative AI and VTubing.

# 2 Digital Faciality Culture and Identity

Digital technologies shape how we think about ourselves and others. Contrary to early beliefs about a fully neutral and democratizing Internet, digital technologies have reified the political boundaries of identity categories along the axes of race, gender, and sexuality, contributing to their formation in a contextual sociocultural consciousness. One arena in particular that has been mediated by the digital is the construction of the face, which can be understood as a technology that "do[es] not come ready made, but [is] produced by an 'abstract machine of *faciality*,'" as Gilles Delueze and Félix Guattari have theorized [14]. That is, faciality is concerned with the making of the face and what beliefs, meanings, and significations we associate with it. Digital faciality culture, then, is the discursive venue in which the digital mediates what the face signifies, such as what aspects of identity we believe can be inferred from the face.

This literature review traces the ways in which digital faciality culture has shaped identity. Of course, the construction and performance of identity in the digital did not start with the face. Before approaching how digital faciality culture has affected identity, it is important to first review foundational work studying how the digital has *been* a space where identity is constructed and played with, starting with the textual.

## 2.1 Playing with Identity from the Textual to the Visual Internet

Before today's multimedia Internet, filled with voice chats, livestreams, and endless feeds of video clips, the Internet was primarily textual. Some of the earliest scholarship on identity in digital spaces focused on this textual Internet, especially with how identity played out in the landscape of BBSs ("bulletin board systems") and MUDs ("multi-user dungeons"), where individuals could log in under pseudonyms and act out an identity of their choosing. Sherry Turkle [57] focused in particular on how these digital spaces helped to reify the multiplicity or distributed nature of identity, and how multiple parallel selves could be played with, where "for some, this play has become as real as what we conventionally think of their lives." Turkle explored various case studies of men who would even play with their gender in these online spaces, such as the user "Case," who played as multiple different women, which Turkle saw as an "experimentation" with a multiple and fluid "self" [58]. Similarly, in Howard Rheingold's early observations on the online community known as "The WELL," he described users such as "Joan," who identified as a disabled woman online, but "who in real life, IRL, was neither disabled, disfigured, mute, nor female" [46]. Though this sort of identity play might have come at the cost of deception, he noted that this technology "dissolved boundaries of identity," in alignment with Turkle's claim that "a more fluid sense of self allows for a greater capacity for acknowledging diversity."

However, it was not necessarily true that identity play in cyberspace allowed for the blurring of identity boundaries and the increased acceptance of a diversity of identities. Through interviews and participant-observation ethnography conducted on the MUD codenamed "BlueSky," social scientist Lori Kendall [27] provided evidence to the contrary against the positions of Turkle and Rheingold, that "despite performing different identities online, many participants continued to believe in essence and continuity of identity," revealing the digital not as a place that necessarily *changed* users' core identity, but rather a place that gave them opportunities to temporarily participate in what they perceived as "a just and egalitarian world" where their real-life ("rl") identities did not matter. Lisa Nakamura [35] conceptualized this sort of brief online participation into "othered" identities as "identity tourism," where the dominantly White population of computer users at the time could transiently perform as identities not their own in textual online spaces. As opposed to the popular belief that the Internet was neutral and color-blind, Nakamura showed how a presumed Whiteness existed in cyberspace even when race was not explicitly mentioned, and when it was, how racial stereotypes were reified by the performances of these "identity tourists." She describes in *Cybertypes* [36] that "like tourists who become convinced that their travels have shown them real 'native' life, these identity tourists often took their virtual experiences as other-gendered and other raced avatars as a kind of lived truth," when in reality they were simply playing out their stereotypes of race and gender. Though one might argue that this sort of identity play is less possible in today's highly visual Internet, recent advances in face-based AI technologies may serve to change that, requiring us to revisit the possibilities and implications of identity tourism.

By the turn of the century, the Internet was no longer purely a textual space; instead, it had grown to become a more visual medium, where images and graphics (including graphical interfaces) abounded. Nakamura comments in *Digitizing Race* [37] that while *Cybertypes* "focused on the constraints inherent in primarily textual interfaces that reified racial categories, … in this work I locate the Internet as a privileged and extremely rich site for the creation and distribution of hegemonic and counterhegemonic visual images of racialized bodies." That is, the Internet, or cyberspace, now aided in the process of identity formation *through* the production and sharing of the visual. In reference to the aforementioned case study of alllookthesame.com, Nakamura [37] argues that "it emphasizes the ways in which the *visual* has always had primacy in our understandings of race." Though it is important to discuss how visual design and interfaces play a role in racial formation on the Internet, the late 1990's and early 2000's introduced new forms of visuality on the Internet, ranging from the use of ASCII art in bulletin board signatures to early uses of JPG and GIF images to serve as visual representations of users, known as avatars. It was not that human society had become suddenly more engaged with the visual, but rather that visuality was now able to be captured and mediated by the digital. Users would be able to create avatars to virtually re-imagine and re-present themselves on forums and instant messengers.

Advances in digital technologies throughout the early 2000's, such as faster Internet speeds, larger storage capacities, and the embedding of cameras to many (if not all) devices brought the visual Internet to an entirely new level, with the face becoming more prominent, whether on video platforms such as YouTube (recalling the dominance of early YouTubers such as Michelle Phan who specifically produced make-up and "DIY face care" content that invoked much conversation about her Vietnamese American identity [56]) or in the new phenomenon of social networking sites, such as Facebook (which itself started as a misogynistic platform for Harvard students to rate women "hot or not" by their faces [22]). By 2013, the face would become so ubiquitous with digital technology use that Oxford Dictionaries would declare "selfies" (or photos of the self, typically of one's face), as its Word of the Year [9]. In reflecting on faciality decades earlier, John Welchman [60] hypothesized, "The face is probably the primary site of visual representation, and has shaped the very conditions of visuality"; certainly, the face today plays a primary role in the visual Internet, and has become a focal point for mediating identity along the axes of race, gender, and sexuality. Whether in reference to selfies, the popularization of face filters, or the automated cropping of images on social media to center faces [66], the face has become a central part of our digital experiences, with many of these technologies only now possible as a result of advances in the subfield of artificial intelligence known as computer vision. Along these lines, Scheuerman, Denton, and Hanna [51] found that of the 487 most popular computer vision datasets, 205 (or 42%) of them were "face-based," pointing to the reality that there is a dominant focus on "facial recognition technologies" (FRTs), which play a crucial role in shaping the very contours of digital faciality culture.

## 2.2 Facial Recognition Technologies and its Critiques

Though FRTs are almost synonymous with computer vision today, these technologies emerged much earlier, originating in the late-1960's from researchers in both Japan and Palo Alto [16]. Kelly Gates traces this early history of FRTs, arguing that the sociocultural beliefs of what FRTs are capable of (whether true or not) have been as instrumental to their construction as much the actual technical realities of these technologies. For example, much of the early work on FRTs was funded by the US Department of Defense, and focused on the detection and identification of the face specifically for "defense" and "national security" applications, which would guide the objectives of FRT research over the decades to come. Following 9/11, FRTs would be praised in the Western world as a technosolutionist tool that could help win the "war on terror" through the identification and tracking of a conceptual "face of terror," [16] a metaphor for the racist visual stereotypes of Middle Eastern, South Asian, and West Asian individuals perceived as "terrorists" at the time.

The history of FRTs tell a story of how race, identity, and the face have always been inextricably intertwined. While Gates has focused on the formation of the "face of terror," Black feminist scholars such as Simone Browne [8] have shown that Blackness has been central to the project of surveillance since before the existence of FRTs, rooted in methods of tracking and policing Black life during slavery, while Safiya Noble [39] has explored the inherent racism and oppression of face-based algorithms such as Google Photos tagging Black individuals as "gorillas" or Nikon cameras incorrectly detecting Asian faces as blinking. This early critical work formed the basis of ongoing threads of scholarship that have critiqued the racist and sexist nature of FRTs, such as the often-cited "Gender Shades" paper from Buolamwini and Gebru [11], which analyzed how commercial facial detection systems identified the faces of dark-skinned women at over 30% worse accuracy compared to the faces of light-skinned men. "Gender Shades" would lead computer scientists (particularly within the "Fairness, Accountability, and Transparency" community) to call for more representative datasets, inclusive of more photos of Black and Brown individuals in order to address these accuracy problems. Writing in 2019, Buolamwini [10] wrote that technologists can "solve" gender and racial bias through "full spectrum inclusion" in datasets, while simultaneously pledging to "prevent the lethal use and mitigate abuse of facial analysis and recognition technology."

However, activists and critical scholars would argue that the increased inclusion of Black and Brown people in image datasets would lead to further capture and surveillance of historically oppressed communities. For example, Tawana Petty, director of the Data Justice Program for the Detroit Community Technology Project, has worked to ensure that marginalized communities are *not* visible to policing systems such as Project Green Light, a city-wide surveillance system deployed by the Detroit Police Department that utilizes FRTs to identify and arrest potential criminals. Petty [42] describes how Detroiters are trading the perception of safety for surveillance, which has actually hindered community safety efforts and directly harmed community members. Perhaps most notable is the case of Robert Williams, who was wrongfully arrested in January 2020 as a result of Project Green Light, possibly the first instance of a wrongful arrest by a facial recognition system [23]. By this time, computer scientists such as Raji et al. [45], including both Buolamwini and Gebru, would shift their focus to the use of "algorithmic audits" to rectify the harmful impacts of FRTs, stating that these technologies "need to have careful privacy considerations, and avoid exploiting marginalized groups in the blind pursuit of increasing representation." Yet, much scholarship today continues to focus on the topics of bias and fairness in machine learning datasets and algorithms.

There is a sense that FRTs have failed across identity lines, in particular in failing to serve people of color, and that this results from dataset bias, or the lack of inclusion of certain identities in computer vision datasets. But such an argument implicitly assumes that there *are* physiological differences between individuals of different identities. Cultural scholar Claudio Celis Bueno [12] reminds us that "the human face should not be considered a universal or natural given but must instead be seen as the product of a specific social assemblage," reminding us that the basic operation of FRTs relies on the mistaken logic that subsets of facial features *do* encode for specific identities, with more representative datasets allowing for more accurate capture of these identities. What we should question instead is whether the task of inferring a person's innate characteristics from the face is possible *at all*. Though there are extremely valid and important implications surrounding surveillance and privacy around the development of facial analysis algorithms, a larger question remains about how the existence and use of such technologies has shifted our cultural understanding of what we believe the face to signify in terms of identity.

## 2.3 The Physiognomy of Facial Analysis Algorithms

Whereas many early FRT applications focused on detection and identification, computer vision researchers in the late-2010's would begin to focus on another aspect of FRTs — that of *facial analysis*, which proposed that computer vision systems could be trained to infer characteristics about an individual from their facial features. Such work includes that of the often-cited Fu, He, and Hou [15], which asserted that it was possible to learn race through the face using a series of state-of-the-art computer vision algorithms. Some of the early (and highly controversial) studies on facial analysis continued in the flavor of surveillance and the criminal legal system, such as Wu and Zhang's "Automated Inference on Criminality using Face Images," where the authors developed a computer vision model that they claimed could achieve 89.51% accuracy in correctly determining if an individual was a criminal based on their face [63]. After an incredible amount of criticism of their study across social media, Wu and Zhang released a "Response to Critiques" commentary article a year later, in which they asserted that such a task of detecting criminality *is* possible, given that "well-trained human experts can strive to ensure the objectivity of the training data" [64]. Many such papers purporting to be able to infer innate identity characteristics from images of the face, such as race, attractiveness, and emotional state, began to appear afterwards, most notably the infamous "Stanford Gaydar" paper by Kosinski and Wang [59], who developed a facial analysis algorithm that they argued could detect if an individual was gay based on an image of their face.

While their most accurate model had an accuracy of 91%, which is worse than a system that classifies *everyone* as straight since 93% of the population identifies as such, they took their finding to mean that it *was* possible to infer sexuality from the face, that such a correlation implied a mechanistic relationship between one's facial features and their identity. Arcas, Mitchell, and Todorov [65] specifically analyze the cases of Wu and Zhang [63] and Kosinski and Wang [59], rooting the logics of these applications in the racist pseudoscience of physiognomy, which asserts that a person's facial features could be used to infer their inner characteristics through correlation. Correlation, of course, does not imply causation. There are no biological mechanisms that provide causal explanations for connecting facial features with innate identity characteristics, beyond the late-19th century speculations of Francis Galton, Karl Pearson, and Cesare Lombroso who employed scientific language to justify their racist beliefs. For this reason, Stark and Huston [55] call these facial analysis technologies "physiognomic AI," or "[t]he practice of using computer software and related systems to infer or create hierarchies of an individual's body composition, protected class status, perceived character, capabilities, and future social outcomes based on their physical or behavioral characteristics." They argue that these technologies present dangers to civil liberties and offer policy recommendations such as wholly banning physiognomic AI from use "at the local, state, and federal levels," even calling upon the FTC to develop consumer protections.

Despite criticisms from scholars such as Stark and Huston [55] and Arcas et al. [65] that the efficacy of physiognomic AI systems is "dubious, unjust, and discriminatory," there still exist applications like that of Faception, an Israeli company whose "main clients are in homeland security and public safety, suggesting that there are surveillance cameras in public places today being used to profile people," sorting them into specific pre-designated categories such as "terrorist," which is reminiscent of the controversial facial analysis system developed in China to identify and track the dominantly-Muslim Uyghur minority group [34]. Though their arguments ultimately utilize the concept of physiognomic AI to argue against mass surveillance once more, their work in tracing the logic of facial analysis algorithms back to physiognomy aids us in understanding how such technologies have influenced digital faciality culture, specifically with how FRTs have renewed and amplified the belief that qualities such as criminality and sexuality can be expressed in the face.

Similar to using facial analysis to detect *sexuality*, a collection of technologies to infer *gender* from the face called "automated gender recognition," or AGR, has also emerged. In one of the first critical pieces of scholarship critiquing AGR, Os Keyes [28] calls these technologies "misgendering machines," drawing from cisheteronormative logics of gender that exclude trans and non-binary identities. Keyes specifically discusses how such technologies operationalize gender, presenting it as an immutable binary classification problem. This, of course, contradicts the reality of gender identities beyond the "male" and "female" binary, as well as the fact that one's gender is not static and immutable, but can be dynamic, as in the case of trans and genderfluid individuals. Through several interviews with trans and non-binary individuals who have been exposed to AGR, Hamidi et al. [19] confirmed that these systems did not work for trans and non-binary individuals, posing safety and privacy risks, and adding to a sense of gender dysphoria. Furthermore, Scheuerman et al. [53] conducted a systematic review of commercial AGR technologies currently used in practice by Amazon, Clarifai, IBM, and Microsoft, evaluating them against a custom dataset of 2450 photos (posted on Instagram and including a gender hashtag) that includes five additional gender identifications beyond "male" and "female." They found that different AGR technologies were unable to label beyond the gender binary (because they were only trained on it) and that different applications often labeled the same photo with different gender labels. They found that how users saw their own identity was not aligned with how computer systems saw their identity, in a way that could cause gender dysphoria while asserting the "truth" of binary gender classifications.

These studies are reminiscent of Kelly Gates's comment that "automated facial recognition systems make use of socially constructed classification systems for defining and standardizing identity categories," [16] which we can understand to be political in nature and developed by institutions for the purpose of control, as Bowker and Star argued in their canonical work *Sorting Things Out* [7]. This also mirrors the analyses presented by early digital scholars as well, such as Nakamura [36] who explored "menu-driven identities," where part of the process of racialization in digital spaces occurs in the selection of race from a set of pre-designated categories. What arises are critiques of the deliberate decisions that are made in determining the identity categories that are used in face-based datasets; Scheuerman et al. [54] find upon reading through documentation for 113 "face-based" datasets that "impartiality" and "universality" are strived for in the creation of these datasets, revealing the belief of dataset creators that an objective "view from nowhere" [20] is possible when analyzing faces. Instead, Scheuerman et al. [54] argue that algorithmic systems developed from these datasets have a particular "way of seeing" that is situated and contextual to how their creators conceptualize gender and other identity categories. Santo [49] builds upon these arguments to show how the categories used in datasets are particularly influenced by political histories, tracing how the inclusion of racial categories has specifically been tied to the US Census categories. The "pan-Asian" racial category, for instance, was explicitly political in origin, but has since been treated as biologically real and essential in the processes of developing face-based datasets.

However, the inferences made by these algorithms are obscured behind what Katherine McKittrick [33] calls "the seeming neutrality of mathematics—the governmental trust in the technologies that calculate the textures of skin, eyes, hair" that causes these algorithms to be treated as objective. This echoes what Ruha Benjamin [4] conceptualizes as "the New Jim Code," which she defines as "the employment of new technologies that reflect and reproduce existing inequities but that are promoted and perceived as more objective," allowing for racism and oppression to hide under technological claims of neutrality or color-blindness. Through the lens of digital faciality culture, these arguments can be understood as meaning that facial analysis algorithms operate in a way such that they give faux legitimacy to a particular situated view of the face, which is taken to be objective. For Scheuerman, Pape, and Hanna [52], the key danger here is the ability for these algorithms to be exported across the world, outside of their typically Western context, asserting the "truth" of these categories through a colonial process that they call "auto-essentialization." Indeed, I argue that the process of auto-essentialization has already shaped digital faciality culture, reinforcing the fallacious belief that the face can signify objectively real identity characteristics.

In 2022, such a phenomenon became a reality with the launch of Giggle, which was marketed by its creators as a "female-only social media app." The catch? Giggle used facial recognition technology to detect if a user was "objectively" a woman or not before allowing them to access the platform. Directly from their website:

> "Bio-metric gender verification software ensures that those within the platform are verified girls. This involves a 3D selfie that performs a quick study of the person's bone structure to determine the female gender. It's science!" [5]

Of course, given the known issues of FRTs and AGR, many women who did not fit a normative image of womanhood were denied access to the app, including many trans women. The creators of Giggle used their FRT/AGR system to reiterate an ongoing anti-trans narrative of biological gender, and the physiognomic logic of being able to understand gender identity through the face.

Though not all believe in the ability for such technologies to accurately detect one's gender, there are still individuals and technologists, both "trans exclusionary radical feminists" (TERFs) and trans individuals alike, who believe that such a task *is* possible, who believe that FRTs are reflective of an objective truth in determining identity. Kit Chokly [29] found in interviews with trans users who have encountered AGR technologies such as Giggle that "I think the most validating experiences can come from artificial intelligence, that you know aren't just gonna say you look xxx to make you feel good." While there may be some who use this technology to specifically defend or enact their essentialist beliefs of race, gender, and sexuality, this technology furthermore has the ability to influence, discipline, or norm others into buying into physiognomic claims of identity in the face. As Chokly [29] concludes, "technologies like AGR are already part of the feedback loops of gender production, for better or for worse."

## 2.4 Playing with the Face

With face-based AI technologies come subversive practices to play with the face in a way that breaks or challenges these technologies, such as in the case of trans users playing with AGR systems to try to "pass." But play with faciality did not merely begin with the digital. It extends far before algorithmic technologies, and even before Deleuze and Guattari's cultural theory of faciality. Histories of play with gender in the face, such as in the classical all-men Japanese theatre form of kabuki and the more-recent all-woman performances of the Takarazuka Revue, have utilized elaborate uses of masks and make-up in order for actors to perform as various genders beyond their own and beyond the gender binary [26].

Within the institutional history of the United States, Wendy Sung [61] has shown how one of the earliest official uses of the face as identification was for the identification of Asian immigrants after the passing of the Chinese Exclusion Act, revealing that "our nation's surveillance system is based on the Chinese face during the Chinese Exclusion Act era." This, famously, led to the "paper sons" or "paper daughters" phenomenon, where Asian immigrants made use of their inscrutability to Western onlookers in the face in order to play with identity and subvert the Chinese Exclusion Act to immigrate into the country [38]. Recalling the arguments of Simone Browne regarding the paradox of Blackness, Sung constructs the Asian face as "sociotechnical," simultaneously hypervisible as immigrants yet individually indistinguishable.

Chinese immigrants were required to complete Form 430, which required an attached photograph to verify identity. As Nham [38] notes, "these identity documentation practices cast identity as an empirical and immutable phenomenon," instead of something dynamic. In other words, facial technologies, including photography and FRTs alike, have been a way to affix a static identity onto the face. Palumbo-Liu [40] discusses how "the face is elaborated as the site of racial negotiations and the transformation of racial identity" as part of the project of nation-making, including defining what citizens of a nation look like. For Asian individuals, our faces are in some ways defined by their inscrutable nature to the White gaze; Sung sees this as a *feature* built into face technologies that allows us the ability to subvert identification. As in the case of "paper sons" and "paper daughters," Asian individuals have utilized play with the face in order to subvert the government and its attempts at forming a "disoriented" facial racial identity that is legible to the White gaze.

Since then, there has been a continued history of attempts to subvert the government, as well as surveillance more broadly. Returning to the digital, CV Dazzle [21] is an early project developed by Adam Harvey in 2010 that he calls "anti-surveillance camouflage," which can be used to subvert the surveillance of FRTs. Specifically, CV Dazzle described a set of possible recommendations for altering one's analog face through the use of makeup, obscuring of the nose and eyes, and alternative hairstyles to block the Viola-Jones face detection algorithm, which is included in the dominantly popular OpenCV computer vision library. In the years that have followed, feminist and queer researchers, practitioners, and artists have launched a variety of projects to subvert the reading of identity from the face, such as Sasha Constanza-Chock's work with the Algorithmic Justice League on "Drag vs AI" [3, 24] to throw off surveillance systems using the queer cultural practice of drag makeup, which acts "as a tool to subvert, fool, and refuse recognition and classification" both historically and in reference to FRTs. Similarly, as a response to the "Stanford Gaydar" algorithm, Zach Blas developed the "Facial Weaponization Suite" [6] to use aggregated biometric data to create physical masks to confuse facial recognition and analysis systems. Taking inspiration from masks for protest as well as the illegibility of the Asian face leading to "biometric failures," Blas draws on the idea of collective protest to create a 3D-printed "collective mask … produced by aggregating the biometric facial data of participants" that participants can wear to cause FRTs to fail in an act of what he calls "defacement," or protest against facial surveillance and the "governmentalities of the face."

But the play and performance of identity via the face go beyond the function of subverting surveillance, as game and media studies scholar PS Berge argues. Berge [5] analyzes the art installation "misgendering machine" (named after the paper of the same title by Os Keyes) developed by the Danish artist ada ada ada, which plays with an automated gender recognition system, showing how the face is dynamic rather than static, and how the dynamic quality of the face can shift how these digital systems recognize our gender. ada ada ada notes that "if you want the machine to gender you as a man, you have to embrace the manly gender stereotypes from the software's data set … of being either angry or without emotion" [2]. Through their project, they show how AGR technologies are "built on gender stereotypes, not a proper understanding of gender" through the ability for users to easily subvert, trick, and/or play with the system by changing visual aspects such as their facial expressions and the angle in which their face is seen by their webcam.

Rather than focusing on how to improve or subvert these systems, and thus buying into the logic that the face can objectively infer identity, the point here is finding *joy* or euphoria in the playing of our identity via our face. As opposed to "passing" in the eyes of the cisheteronormative and physiognomic misgendering machines, PS Berge [5] argues that trans individuals are "finding play in being unrecognizable and in unraveling malicious fantasies of authenticity, verification, and computational ontology." It is not that we are buying into the essentialist narrative of physiognomic AI, but rather that we get to play with and learn to subvert the cisheteronormative understanding of gender through its reproduction by the AI technology. Similarly, in response to "gender swap" face filters that have become popular on apps such as Snapchat and TikTok that have reinforced the gender binary and led to trans users experiencing gender dysphoria, Kat Brewster [25] has conducted speculative design workshops with trans individuals to design a face filter application that would allow for users to play with aspects of their face and envision alternative, affirming realities, without making specific reference to gender categories and the facial stereotypes associated with those genders, while protecting their privacy and giving users control over how their data is used and who it can be seen by. These examples offer us insights into how scholars and artists have been able to utilize the digital in order to play with identity in a time when facial recognition and automated gender recognition technologies abound.

## 3 Digital "Post-Faciality" and *Trans*gressions

Face-based AI technologies such as FRTs and AGR have served as a particular discursive and material context in which the relationship between identity and the face are strengthened. These technologies have shaped the contours of digital faciality culture through an attempt to add legitimacy to the logics of physiognomy and the perceived reality that politically defined identity categories can be directly inferred from facial features. However, FRTs and AGR are not the only face-based AI technologies that exist today. Drawing from the exact same datasets and logics of FRTs, new face-based AI technologies such as generative AI, deepfakes, and VTubing (which I conceptualize as a form of animated deepfake) have emerged in the previous few years. I call these "post-facial" technologies, which are still rooted in the primacy of the face but simultaneously allow us to eschew visual ties to our analog faces. In this section, I will introduce my conception of the "post-facial" and argue that much work remains to be done on studying how these technologies continue to complicate our relationship (or offer possibilities) between identity and the face in the digital.

### 3.1 Post-Facial Technologies

Whereas technologies such as face filters rely on digitally modulating aspects of a user's analog face, post-facial technologies, which include — but are not limited to — deepfakes, VTubing, generative AI faces, Memoji, and Apple Vision Pro Persona, need no visual relationship to one's analog face at all. For example, deepfakes utilize facial recognition technologies to map a wholly different face onto an existing one, emulating facial expressions and movement while eschewing the actual features of the original face (e.g., skin tone, eye color, mouth shape, etc.). VTubing and Memoji are extensions of this technology, where an animated face (whether in the case of a VTuber avatar or an Apple-branded 3D emoji) replaces one's face in the digital. Apple Vision Pro Persona reproduces a synthetic face that one performs with when using FaceTime while wearing the Apple Vision Pro headset. While such technologies still center the face, both in the presentation of a simulated face and the control of that face using one's own, the actual visual characteristics of one's analog face are hidden.

Just as post-humanism is not a complete departure from the human, the "post-facial" is not a complete departure from faciality. These technologies are still rooted in the same face-based data, draw from the same logics of physiognomy that affix identity characteristics to specific facial features, and are even powered by the same underlying facial recognition algorithms. However, with post-facial technologies, users now have the ability to generate, control, and perform as a face that is not their own. One can imagine that in terms of a simplified Hegelian dialectic, the post-facial can be construed as the synthesis between the previous eras of identification through the face and the responses that play with, resist, and/or subvert facial recognition. No longer are we tied to technologies related to representations of our own faces; rather, we can perform with faces completely separate from our own, ones that signify a different set of identity characteristics.

What is particularly distinct about this class of technologies is that they allow for the face to become the new mode in which we inter*face* with the digital. Whereas in previous eras of the Internet, "users can express their subjectivity while mouse clicking their way through the web," [36] we are now at a moment where we can directly interact with our technologies using our face. Instead of typing out our passwords, FaceID uses our faces to unlock our phones. Countries like China have led in the development of payment systems that directly use facial recognition [31]. Games such as *Before Your Eyes* are controlled by facial expressions and blinking [30]. When analyzing the textual Internet, Nakamura [36] asked: "what structures of organization are mediating users' experiences with the web?" The answer today is that the face as assemblage now mediates our experiences with the digital. A particularly revealing example of a post-facial technology is that of VTubing, which is short for "Virtual YouTubing." It is an extension of online streaming, with its key distinction being the use of a CGI avatar in place of the analog streamer. But as opposed to using a static image to represent the individual, as was the case in earlier online spaces, VTubing is unique for its motion capture technology rooted in computer vision, where the virtual avatar is controlled by an actual person whose motion and facial expressions are reflected and replicated in real-time. Though digital avatars have been popular throughout the visual age of the Internet, whether with customizable avatar generators such as Picrew, or playable 3D avatars in *Second Life*, what sets aside this new post-facial technology is how a user is able to inter*face* with the digital directly through the face. Rather than clicking a mouse or hitting a key to change the expressions on the VTuber avatar, all a user needs to do is directly change their facial expression, which allows for a far more embodied experience. In this sense, the face is still central as the key mechanism for interfacing with post-facial technologies, but the visuality of our own analog face is no longer centered as the object of scrutiny in the digital world.

With the inception and popularization of post-facial technologies, there remain many open questions for the formation and performance of identity. Prior work on identity formation and performance in the digital, complemented by critiques of FRTs based in the logics of physiognomy, can serve as a foundation for the sorts of analyses and theorizing to come. One particular area where this applies is with the post-facial technology of generative AI faces.

While VTubing and deepfakes allow us to perform with faces that are not our own, generative AI allows us to *create* faces that are not our own, raising a related but different set of implications for identity. In their talk on "Automatic Gender 'Recognition' and AI Image Generators," CQ Quinan [44] gives the example of asking DALL-E 2 to generate an image of a "transgender woman" and a "transgender man," both of which fail to produce meaningful representations of trans individuals, but rather reproduces visual stereotypes of how cisheteronormative society perceives of trans individuals. Again, we are confronted by questions of how identity is conceptualized by AI technologies, and how such technologies might possibly shape future perceptions of these identities.

Early work that has attempted to critique such generative AI technologies with respect to identity falls back on earlier narratives of dataset bias and fairness. In "This Black Woman Does Not Exist," Eryk Salvaggio [48] criticizes the failure of StyleGAN, a generative AI model built by NVIDIA, in its inability to produce "convincing faces of Black women." Salvaggio chalks this up to a clear lack of Black women in the public Flickr dataset which StyleGAN is trained on — he notes that "of the 4000 images sampled, 102 contained Black women, or just about 2.55%. By comparison, there were 1152 white women, or 28.8%," definitive evidence of dataset bias. Furthermore, from my literature review, I found that the *only* peer-reviewed paper concerned with questions of identity and AI-generated faces, published in June 2024 by Ye Sul Park, also critiques generative AI solely from the perspective of bias, arguing that models such as DALL-E 2 are "mostly trained on U.S. based data … centering Western epistemologies … reifying White supremacy" [41]. While this is certainly true, Park cites Salvaggio and reiterates the belief that increasing the "racial diversity" of the data set on which the generative AI model is trained serves as an adequate intervention to the visual reproduction of Western stereotypes and norms, particularly with Black and Asian subjects, by these generative models. However, I argue that just as these critiques from the perspective of fairness and bias have been inadequate for analyzing the harms of FRTs, they are just as inadequate for analyzing post-facial technologies.

One can take the case study of Meta's AI image generator, which tech journalist Mia Sato has shown cannot generate an image of an Asian man with a White woman [50]. Though Sato entered various prompts to the AI generator, including "Asian man and Caucasian woman couple" and "Asian man with White wife," it continued to only generate images of Asian men with Asian women, whereas prompts like "Asian woman with Caucasian husband" or "Asian woman with African American friend" worked. Also notable is the fact that all of the generated images of Asians depicted stereotypically light-skinned individuals, failing to capture the vast diversity of appearances of "Asian" individuals. From one perspective, while there may be a sense that larger datasets with more diverse and representative data may address one aspect of the issue, I argue that we must think more deeply about how generative AI models "see" identity. When asking a generative AI model to generate an "Asian" face, how and what does it conceptualize as "Asian"? In particular, the example of generating an Asian man and White woman is perplexing to traditional arguments of biased datasets, because image datasets *do* contain ample amounts of data of both Asian men and White women, and generative AI systems *are* able to construct faces of both independently. For this reason, we need to think about how identity is constructed by generative AI technologies beyond the explanation of biased datasets, and how this in turn influences how we think about identity categories such as race and gender as represented by the face. I argue that there is a gap in our scholarship for thinking about the impact of generative AI and other post-facial technologies from a more critical and sociocultural perspective. This paper does not offer definitive answers to fill the gap, but rather proposes some directions in which to proceed by drawing on existing scholarship and practice.

## 3.2 Playing with Identity with Post-Facial Technologies

One particular avenue of interest to me is how identity can be played with through the use of post-facial technologies. We can read previous scholarship on the multiplicity and performance of identity from Turkle, Kendall, and Nakamura anew in this post-facial age. For example, because VTubing and deepfakes allow for individuals to perform faces that are completely distinct from their own, I argue that this allows for a return to the possibility of identity tourism once more. Whereas previous facial technologies such as facial recognition worked to affix identity onto facial characteristics, and individuals were rendered visible on the visual Internet by way of webcams, post-facial technologies allow us to navigate digital spaces while possibly embodying wholly different identities, just as the textual did.

But just as with FRTs, we still run into the same logics of physiognomy with post-facial technologies. If we consider that we can perform specific identities through deepfakes or construct specific identities through generative AI, this is still rooted in the belief that identity and the face are still inextricably connected, that the face *does* correlate or correspond to innate identity characteristics. Though we might be subverting the primacy of the analog face, we still place value in how the face informs our perceptions about identity. As post-facial technologies are not simply a complete departure from a culture of faciality, it can be argued that they are still technologies built from physiognomic logics. Perhaps just as Audre Lorde [32] said, this is a case where "the master's tools will never dismantle the master's house," that though post-facial technologies may challenge, disrupt, or complicate how identity is constructed in the digital, it is still not liberatory enough.

However, this is not to say that post-facial technologies do not have *any* influence on how we come to understand, perform, and form identity in the digital. For those who utilize post-facial technologies, I am curious about the lived experience that comes with it, such as the possible joy that trans VTubers gain in being perceived as a gender that is more in line with their own. At the same time, we return to the same complications of identity tourism as described by Nakamura in the textual internet, specifically that of racial fetishization. Because VTubing draws from Japanese aesthetics and "kawaii" culture, it has raised the question of digital yellowface, an emerging term that first appeared in response to a variety of Snapchat filters that gave users stereotypical and/or caricatured Asian facial features. With VTubing, the race of the streamer cannot be visually inferred, since the virtual avatar acts as proxy for the appearance of the performer. As opposed to the textual internet where Whiteness is seen as default, Asianness is the assumed default in the landscape of VTubing. Behind the anonymity of their virtual avatar and the assumed Asianness, some VTubers have adopted stereotypically Japanese pseudonyms (with backlash from fans and peers), often performing a form of Asianness through the inclusion of mimicked Japanese phrases (e.g., "Konichiwa Bitches" and ending sentences in "-desu"), even when it is later revealed that these streamers are White and not in fact Japanese. This is reminiscent of Nakamura's analysis of the identity tourism, where "this type of Orientalized theatricality is a form of identity tourism; players who choose to perform this type of racial play are almost always white, and their appropriation of stereotyped male Asiatic samurai figures allows them to indulge in a dream of crossing over racial boundaries temporarily and recreationally" [35]. With VTubing, identity tourism continues to be coupled with the hypersexualization of Asian women and the use of Asianness as nothing more than an aesthetic.

But while post-facial technologies are complicated by histories of yellowface (and blackface) extended into the digital, they simultaneously allow for the possibility of playing with identity in a way that may be liberatory for some, such as trans users. VTubing allows for transformative possibilities in terms of gender expression by trans, non-binary, and queer streamers via the exact same mechanism of proxied visual identity resulting from the virtual avatar. In the following subsection, I connect these discussions of post-facial technologies with trans studies and practices, which have historically grappled with playing with identity via the face in both the analog and digital.

### 3.3 Trans Technologies and Transgressions

> *"Transness is the movement of life across social categories … transness is a practice of freedom."* — Susan Stryker, at the 2nd International Trans Studies Conference

In the way that information science scholarship such as Bowker and Star [7] provide us critical tools to think about the oppressive force of classification, trans studies taken broadly navigate us towards breaking, subverting, or transgressing the boundaries of these classifications. Within digital faciality culture, trans practices have worked to challenge or complicate the project of classification from technologies of faciality, such as physiognomic AI. This subversion is not new. It is no coincidence that many of the studies and projects that challenge the culture of an "objective" faciality, reviewed throughout this paper, have been produced by queer/trans scholars and artists. As scholars such as Legacy Russell and Whit Pow argue, trans folks have always existed as "glitches" to these systems of capture and classification, and this characteristic of the glitch has given us the opportunity for liberation [43, 47]. Glitches can be understood as a "disruptive and deconstructive act within a computational system," one that interrupts how the digital mediates classification, visuality, and embodiment. Along these lines, post-facial technologies *can* be utilized as an extension of queer and trans embodiment, one that transgresses against the identity categories reified by FRTs. Perhaps the post-facial can even become a transgression of faciality itself, challenging the very aspects of identity that we attempt to infer from the face.

Since post-facial technologies such as VTubing and deepfakes allow for play with one's identity, they can perhaps be considered "trans technologies," a term coined by Oliver Haimson [18], who identifies technologies that allow for "the queer aspects of multiplicity, fluidity, and ambiguity needed for transition." Just as with early Tumblr, post-facial technologies such as VTubing — and deepfakes in general — allow trans users to "safely experiment and explore their gender expression." These post-facial technologies offer affordances of anonymity that function not only for privacy to protect the safety of marginalized users, but also to help obscure identity characteristics that may be inferred from one's analog face. Individuals can be openly authentic — which Haimson calls "identity realness" — to themselves, without fear of their real-life identity being compromised. Post-facial technologies can reflect alternate presents or future imaginaries of what the users look like, accounting for a visual identity that can and does change over time. We can use these technologies to reflect what we imagine we look like, what we hope to look like, or simply to challenge the conventions of gender and identity performance. In this way, post-facial technologies *can* serve as trans technologies, "enabling users to change over time within a network of similar others, separate from their network of existing connections, and to embody (in a digital space) identities that would eventually become material" [18]. For example, after interviewing the trans VTuber LemmaEOF, who uses a robot avatar, Haimson [17] writes, "I see endless potential for creating new trans possibilities and customizing self-presentation in inventive new ways." Through this post-facial technology, "LemmaEOF can represent its multiple marginalized identities—trans, autistic, and fat—in the form of a robot, increasing representation for each of these identities and their intersections and helping it feel comfortable while streaming."

But here, we must avoid falling into the same issue that early scholars of the textual Internet did — as Nakamura [36] reminds us, "Piles of articles have been written about cross-gender passing, or 'computer cross-dressing,' but very little has been done on the topic of cross-racial passing." We must remember that trans studies must not simply be concerned about gender, but about race as well. We must simultaneously think about the complex implications that these technologies have for transgressing race, as much as we focus on their liberatory possibilities for transgressing gender. For this reason, it is not to say that post-facial technologies are always in service of justice or resistance, or even trans interests, but rather, that they simply have the potential to be transgressive to the project of classification and identification. Though they may allow us the renewed ability to play with identity, they can present dangers in the realms of consent, disinformation, and the reification of identity stereotypes as a result of identity tourism. Trans scholars and practitioners have given us tools, concepts, and practices to play with our digital interfaces, and similarly, we can view post-facial technologies as having the potential to transgress the boundaries and impositions of the previous era of visuality. But without a framework or practice of transgressing faciality itself, we can just as easily continue to use these technologies to carry on the racist logics of physiognomy that ascribe the face to specific identity characteristics.

# 4 Proposed Study: Interviewing VTubers

To examine the impact of face-based "post-facial" AI technologies and the possibility of transgressions, I propose to study VTubers, who currently utilize post-facial technologies to perform particular instantiations of identity in the digital. As VTubing has become an increasingly popular modality for online streamers, especially queer and trans streamers of color, it should no longer be considered a niche digital practice. For example, the disabled Latina American VTuber Ironmouse became the most subscribed account on the streaming platform Twitch in February 2022, notably becoming the first female streamer to gain that status. In September 2024, Ironmouse would once again reclaim the title as the most subscribed Twitch streamer, reinvigorating conversations around the popularity and affordances of VTubing for traditionally marginalized streamers.

Despite the prominence and cultural impact of VTubing today, nearly no attention has been given to VTubers in digital and media studies, even by those studying streamer culture. This study is motivated by a desire to better understand how VTubers conceptualize their identity as they use this technology, and what affordances it provides in performing identity.

Just as the textual web allowed trans individuals to fully be seen in the way they desired, the affordances of VTubing allow for safer, liberatory possibilities in the current era of online visuality and digital faciality culture. For disabled VTubers like Ironmouse, it allows them to present in ways that can bring attention to their disabilities, while giving them an accessible technology that materially enables them to stream. It is also important to recognize that VTubing is an international phenomenon, not localized to Japan nor the United States, with countries such as Indonesia having thriving VTuber communities as well. Given the pseudo/anonymity of VTubing, Indonesian streamers such as Hana Macchia are able to more safely present their queer identities under the protection of VTubing, especially against the backdrop of growing anti-queer sentiments in Indonesia [62], without fear of backlash or harm to their real, analog selves. Furthermore, VTubing has also allowed for streamers to better embody non-binary and even non-human identities, with VTubers such as Ying (莺) designing their virtual avatar to specifically appear androgynous and Fuwa performing with a frog avatar.

However, the popularly Japanese or anime aesthetic of VTuber avatars cannot be ignored. While VTubing might enable the transgression of gender categories and the inference of gender from the face, the default Asianness and possible aesthetic fetishism of VTubing must be scrutinized. In using this post-facial technology, it is pertinent to inspect how VTubers conceptualize race as well as gender and sexuality as it is performed through the face. While VTubers like Artemis the Blue play with their gender using this technology, questions remain about how the eschewing of the face simultaneously allows them to embody a racial identity that they might not hold beyond the digital. Because of the multifaceted and nuanced nature of VTubing, it serves as a potent object of study for an intersectional analysis of digital faciality culture as mediated by face-based AI technologies. For these reasons, I am curious about how the technology of VTubing has shifted streamers' perception of what aspects of identity we can infer from the face, and how this play and performance informs the experiences of VTubers (if at all).

## 4.1 Proposed Research Methods

In order to carry out this study, I propose the empirical qualitative approach of conducting semi-structured interviews, each an hour in duration, with roughly 20-25 VTuber participants. The goal of these semi-structured interviews is to understand the various experiences of these VTubers, particularly in reference to how they conceptualize and perform their identity through the use of VTubing as opposed to analog streaming. The VTubers who will be interview participants will be recruited through a purposive sample from my (quite large) existing network of VTubers, as well as using snowball sampling based on recommendations from those who have been interviewed. Interviews will then be analyzed using thematic analysis, with iterative coding through the development of a codebook with another coder in order to achieve adequate intercoder reliability.

The goal is to study how VTubers construct and perform with the face, as a particular case study thinking about digital faciality culture today. While VTubing provides room for identity play and varied embodied performance, which may be liberatory in terms of playing with gender, there may be further complications with playing with race, even when unintentional, especially as VTubing has already been implicated with cultural appropriation and aesthetic fetishism due to its roots in Japanese culture. By taking on the general population of VTubers more broadly, as opposed to starting with a specific subpopulation of VTubers, I treat this as an exploratory study that can lead to future directions of research that may not be conspicuous just yet. I hope to look at a broad spectrum of possible topics, including perceptions of gender and race, privacy and safety concerns, and elements of play and joy for a diverse set of participants.

Possible interview questions might include: Why stream with a VTuber avatar rather than with the analog face? What do you think about the identity that your VTuber model conveys to audiences? Has VTubing allowed you to embody a digital identity that you might perceive as distinct from your analog identity?

My hope is that this work may serve as a foundation for future research with VTubers, which would focus on specific facets of VTubing in different contexts. For example, this study would supplement ongoing work focusing specifically on trans VTubers led by Hibby Thach that I am supporting on, analyzing the liberatory possibilities and affordances of VTubing for trans streamers. Finally, this exploratory work may guide future understanding for how post-facial technologies in general might shift our perceptions and performances of faciality in relation to race, gender, sexuality, and disability.

# 5 Conclusion

We exist within a culture of face-making, both literally and symbolically. Within the conceptual framework of faciality, we must come to understand that "faces have to be made and that not all societies require making faces" [12]; in extension, digital faciality culture is the site in which the digital mediates the "abstract machine of faciality" in the process of face-making. Face-based artificial intelligence technologies did not emerge on their own, removed from the sociopolitical context within which they materially and discursively exist. They are products of digital faciality culture, rooted in the physiognomic belief that faces signify essential qualities of the bodies to which they belong. At the same time, they influence our belief that sophisticated enough technologies can accurately reduce identity to a particular subset of facial features, forming a feedback loop to inform our digital faciality culture. It is in this sense that face-based artificial intelligence technologies play a crucial role in the process of both face-making and identity formation vis-à-vis the face.

This literature review traces the instances in which face-based artificial intelligence technologies have interacted with digital faciality culture. Facial recognition technologies, including facial analysis and automatic gender recognition, have been developed in attempts to infer race, gender, and sexuality from the face, yet have largely been analyzed with respect to questions of surveillance, bias, and fairness. Post-facial technologies, such as deepfakes and face generation, continue to be analyzed within the same lenses. As such, we have yet to contend seriously with the question of faciality in the era of post-facial technologies, beyond the notion that the logics that drive these technologies may lie in racist, pseudoscientific origins. Yet, as seen with trans users' engagement with automated gender recognition systems and VTubing, face-based artificial intelligence technologies have the potential to encourage play with the face and shift notions of what a face signifies.

In this paper, I propose *one* possible framework to understand this play as a sort of transgression of the categorizing logics of the past. Post-facial technologies have reshaped the landscape of digital faciality culture by allowing us to generate faces that are not our own, and embody them in the digital as if they were. Will we as a society come to understand identity with respect to the face differently in the years to come as a result of these post-facial technologies and their possibilities for transgression? Or will we simply continue to reproduce the logics of physiognomy and visual subjugation along the same old contours of power and oppression? While inspecting the possibly transformative experiences of those who practice VTubing, there remain so many more applications of post-facial technologies to consider and study. One such example is the role that they play in media beyond the Internet, such as in television and film where deepfakes revive actors of the past in franchises like *Star Wars* and *Alien*. Another dimension to consider is how post-facial technologies can possibly transform the nature of labor, especially at a time when corporations and public services alike have turned to introducing VTuber mascots to personify their brands. This is not to mention the impact that post-facial technologies have on sex work and the visuality of adult content, which has not only been a central focus when discussing deepfakes, but also one of the earliest use cases of VTubing.

Though I began with questions of identity as tied to face-based artificial intelligence technologies, I now find myself questioning the sorts of transgressive possibilities that these technologies can offer us, across the boundaries of race, gender, sexuality, disability, and ultimately faciality itself. For in this era of post-facial technologies, I repeat the question: What's in a face?

To which I answer: Possibility.

# Acknowledgments

Though an acknowledgements section might be atypical for a field prelim paper, I want to emphasize that I could not have made it here without the support of those who I am lucky to be in community with. First, I want to thank my advisors, Matthew Bui and Sarita Schoenebeck, for all their patience with me through the many months to develop this paper, and for giving me their grace and their time to help refine my conceptualizations of digital faciality and provide feedback on paper drafts. I also want to thank my committee members, Lisa Nakamura and Oliver Haimson, for providing me with the intellectual space to grow, whether at conferences, in class, or from their incredible scholarship, which this work builds off of. (Seriously, I could probably write an entire version of this paper that emulates *Digitizing Race* chapter-by-chapter but instead focusing on faciality, and I am extremely thankful for the framework of *Trans Technologies* which has offered me a direction to move my thinking in.) Additionally, I'm indebted to the work of Wendy Sung and Morgan Scheuerman, not just because of their scholarship on digital faciality and its relationship with race, gender, and sexuality, but also because of their support as fellow scholars who have encouraged me to continue with this work. I would not be here without my UMSI cohort, the so-called "si-borgean" group, who came to my aid emotionally more times than I can count, who gave me the momentum and pep talks to stay in the PhD program. Finally, and most importantly, I give endless thanks and love to my partner Amy Wen, for keeping me afloat from afar, sitting with me virtually as we co-worked, and for bringing me joy and a sense of safety in this life. It truly takes a village. I never want to take that for granted.

4, 2024 from https://www.cnn.com/2013/11/19/living/selfie-word-of-the-year/index.html

[10] Joy Buolamwini. 2019. Artificial Intelligence Has a Problem With Gender and Racial Bias. Here's How to Solve It. *TIME*. Retrieved from https://time.com/5520558/artificial-intelligence-racial-gender-bias/

[11] Joy Buolamwini and Timnit Gebru. 2018. Gender Shades: Intersectional Accuracy Disparities in Commercial Gender Classification. 2018. . Retrieved from https://facctconference.org/2018/program.html

[12] Claudio Celis Bueno. 2020. The Face Revisited: Using Deleuze and Guattari to Explore the Politics of Algorithmic Face Recognition. *Theory Cult. Soc.* 37, 1 (January 2020), 73–91. https://doi.org/10.1177/0263276419867752

[13] Haoxuan Chen, Yiran Deng, and Shuying Zhang. 2016. Where am I from? --- East Asian Ethnicity Classification from Facial Recognition.

[14] Gilles Deleuze and Félix Guattari. 1987. *A thousand plateaus: capitalism and schizophrenia*. University of Minnesota Press, Minneapolis.

[15] Siyao Fu, Haibo He, and Zeng-Guang Hou. 2014. Learning Race from Face: A Survey. *IEEE Trans. Pattern Anal. Mach. Intell.* 36, 12 (December 2014), 2483–2509. https://doi.org/10.1109/TPAMI.2014.2321570

[16] Kelly Gates. 2011. *Our biometric future: facial recognition technology and the culture of surveillance*. New York University Press, New York.

[17] Oliver Haimson. 2025. *Trans technologies*. The MIT Press, Cambridge, Massachusetts.

[18] Oliver L. Haimson, Avery Dame-Griff, Elias Capello, and Zahari Richter. 2021. Tumblr was a trans technology: the meaning, importance, history, and future of trans technologies. *Fem. Media Stud.* 21, 3 (April 2021), 345–361. https://doi.org/10.1080/14680777.2019.1678505

[19] Foad Hamidi, Morgan Klaus Scheuerman, and Stacy M. Branham. 2018. Gender Recognition or Gender Reductionism?: The Social Implications of Embedded Gender Recognition Systems. In *Proceedings of the 2018 CHI Conference on Human Factors in Computing Systems*, April 19, 2018. ACM, Montreal QC Canada, 1–13. https://doi.org/10.1145/3173574.3173582

[20] Donna Haraway. 1988. Situated Knowledges: The Science Question in Feminism and the Privilege of Partial Perspective. *Fem. Stud.* 14, 3 (1988), 575–599. https://doi.org/10.2307/3178066

[21] Adam Harvey. 2010. CV Dazzle. Retrieved from cvdazzle.com

[22] Mar Hicks. 2021. When Did the Fire Start? In *Your Computer Is on Fire*, Thomas S. Mullaney, Benjamin Peters, Mar Hicks and Kavita Philip (eds.). The MIT Press, 11–26. https://doi.org/10.7551/mitpress/10993.003.0003

[23] Kashmir Hill. 2020. Wrongfully Accused by an Algorithm. *The New York Times*. Retrieved October 3, 2023 from https://www.nytimes.com/2020/06/24/technology/facial-recognition-arrest.html

[24] Janet Ruppert, Ricarose Roque, and R. Benjamin Shapiro. 2022. Opportunities and Challenges for Enacting Equity and Justice-centered CS Learning in "Drag vs. AI" Workshops. In *Proceedings of the 16th International Conference of the Learning Sciences-ICLS 2022*, 2022. 2052–2053.

[25] Kat Brewster, Aloe DeGuia, Denny Starks, Mel Monier, Ria Khan, Samuel Mayworm, and Oliver Haimson. 2024. Facing the Filters: Designing augmented reality technologies for identity exploration, gender affirmation, and radical possibility by and for the trans

*"The scramble suit was an invention of the Bell laboratories, conjured up by accident by an employee named S. A. Powers... Basically, his design consisted of a multifaceted quartz lens hooked up to a million and a half physiognomic fraction-representations of various people: men and women, children, with every variant encoded and then projected outward in all directions equally onto a superthin shroudlike membrane large enough to fit around an average human.*

*As the computer looped through its banks, it projected every conceivable eye color, hair color, shape and type of nose, formation of teeth, configuration of facial bone structure - the entire shroudlike membrane took on whatever physical characteristics were projected at any nanosecond, then switched to the next...*

*In any case, the wearer of a scramble suit was Everyman and in every combination (up to combinations of a million and a half sub-bits) during the course of each hour. Hence, any description of him - or her - was meaningless."*
— Philip K. Dick, *A Scanner Darkly* (1977)